\documentclass[10pt, conference, a4paper]{IEEEtran} 

\usepackage{cite}
\usepackage{amsmath}
\ifCLASSINFOpdf
  \usepackage[pdftex]{graphicx}
  \DeclareGraphicsExtensions{.pdf,.jpeg,.png}
\else
  \usepackage{graphicx}
\fi

\usepackage{listings}
\IEEEoverridecommandlockouts
\begin{document}
%
\title{SCTP User Message Interleaving\\Integration and Validation}

\author{\IEEEauthorblockN{Felix Weinrank, Irene R\"ungeler, Michael T\"uxen}
\thanks{Parts of this work have been funded by the German Research Foundation (Deutsche Forschungsgemeinschaft).}
\IEEEauthorblockA{M\"unster University of Applied Sciences\\
Dept. of Electrical Engineering and Computer Science\\
Bismarckstrasse 11, 48565 Steinfurt, Germany\\
\{weinrank, i.ruengeler, tuexen\}@fh-muenster.de}
\and
\IEEEauthorblockN{Erwin P. Rathgeb}
\IEEEauthorblockA{University of Duisburg-Essen\\
Institute for Experimental Mathematics\\
Ellernstrasse 29, 45326 Essen, Germany\\
erwin.rathgeb@iem.uni-due.de}}

\maketitle

\begin{abstract}
The Stream Control Transmission Protocol~(SCTP) is a connection and message oriented transport protocol. 
It supports multiple uni-directional streams in each direction allowing user message sequence preservation within each stream.
This minimizes the re-sequencing delay at the receiver side in case of message loss.
The base protocol, although being optimized for small messages, supports arbitrary large user messages by using fragmentation and reassembly at the cost of adding delays at the sender side.
To overcome this limitation, a protocol extension called \emph{User Message Interleaving} is currently being specified by the Internet Engineering Task Force~(IETF).
This paper describes the new extension, its integration and validation in the context of the INET framework.
\end{abstract}
\IEEEpeerreviewmaketitle

\section{Introduction}
\label{sec:introduction}
\begin{figure}[!ht]
\centering
\includegraphics[trim=15mm 100mm 45mm 15mm, clip,width=3.5in]{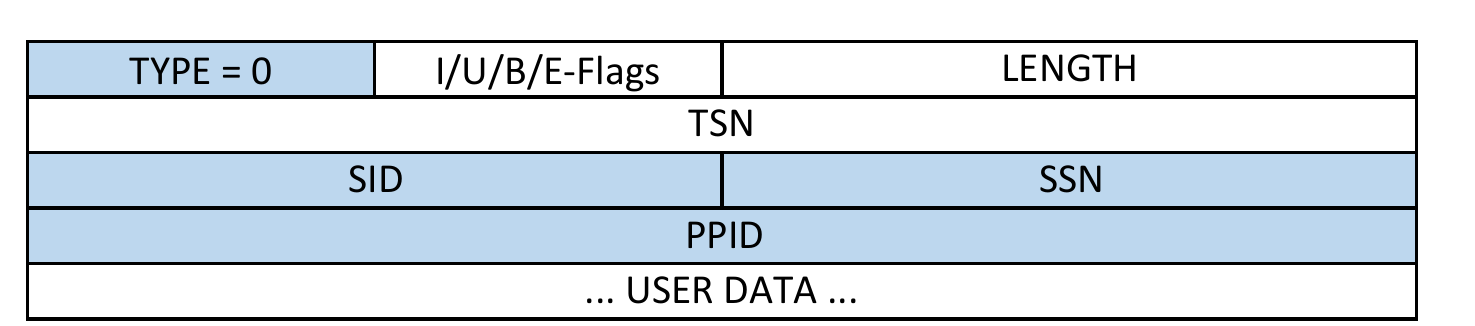}
\caption{DATA chunk structure}
\label{fig:chunk-data}
\end{figure}

SCTP~\cite{rfc4960} is a message oriented protocol optimized for small messages with a special emphasis on network fault tolerance.
In particular, it minimizes the receiver side head-of-line-blocking by supporting multiple uni-directional streams in both directions.
The sender of a user message specifies the stream being used, and sequence preservation is only guaranteed for messages sent on the same stream.
Figure~\ref{fig:chunk-data} shows the DATA-chunk used for transmitting user messages.
Each DATA chunk can be identified by a transmission sequence number~(TSN).
Selective Acknowledgement (SACK) chunks are used to acknowledge the receipt of DATA chunks by referring to the TSNs.
In case of message loss, the sender retransmits the lost DATA chunks.
Therefore the TSNs are used to ensure that the receiver gets all DATA chunks sent by the sender.
The stream identifier (SID) specifies the stream the user message is sent on, and the stream sequence number~(SSN) is used to provide sequence preservation of user messages within each stream.

For supporting user messages larger than a packet, SCTP supports fragmentation and reassembly. 
The sender fragments the user message using multiple DATA chunks for transmission. 
In the first DATA chunk the B-bit is set in the flags field, in the last one the E-bit is set.
All DATA chunks belonging to the same user message get the identical SID and SSN. 
The TSNs are chosen in a consecutive way.
Therefore the TSN is not only used to provide reliability but also to encode the sequence of the fragments.

In most SCTP implementations, the TSNs encode the sequence in which the DATA chunks are put on the wire.
Therefore, if the sender starts sending a large user message, user messages of all other streams are blocked until the sending of the large message has been completed.
This introduces a sender side head-of-line-blocking.

The support of arbitrary large user messages became important in the context of WebRTC. 
SCTP is used as the transport protocol for WebRTC data channels, see \cite{draft-webrtc-dc}. Multiple independent data channels are multiplexed over a single peer connection. 
This is done by mapping data channels on SCTP streams and having a single SCTP association for a WebRTC peer connection.
To avoid the sender side head-of-line-blocking introduced by fragmentation, the SCTP protocol extension to support user message interleaving~\cite{draft-ndata} is specified by the IETF and required for using SCTP in the WebRTC context.

This paper is structured as follows:
Section~\ref{sec:message-interleaving} introduces the user message interleaving extension including the new chunk types.
Section~\ref{sec:stream-scheduler} covers the required modifications to the existing stream schedulers in the SCTP simulation model.
Finally, Section~\ref{sec:validation} show the techniques used to validate the implementation.

\section{Message Interleaving}
\label{sec:message-interleaving}
\begin{figure}[!ht]
\centering
\includegraphics[trim=15mm 95mm 45mm 15mm, clip,width=3.5in]{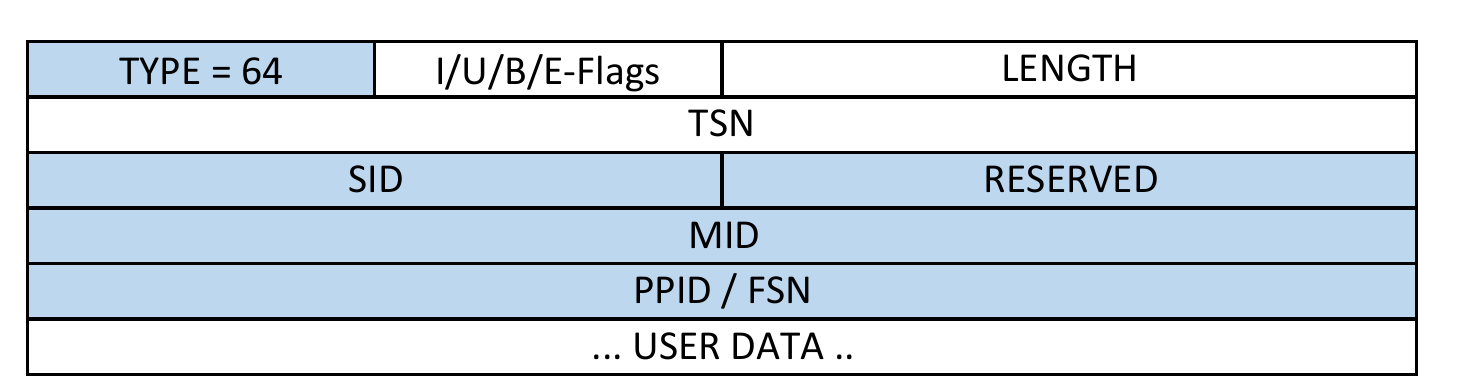}
\caption{I-DATA chunk structure}
\label{fig:chunk-idata}
\end{figure}

To be able to avoid the sender side head-of-line-blocking, user message interleaving uses I-DATA chunks instead of DATA chunks. 
When using I-DATA chunks, the TSN is only used for providing reliability. 
For enumerating the fragments of a large user message, the fragments sequence number~(FSN) is used. 
For avoiding additional overhead, a single field is used to hold the payload protocol identifier~(PPID) for the first fragment, when the B-bit is set.
The FSN is considered 0 implicitly. 
For all other fragments, the FSN is set in that field.
To avoid performance limitations, the 16-bit SSN is replaced by a 32-bit message identifier~(MID).
The I-DATA chunk is shown in Figure~\ref{fig:chunk-idata}.

\begin{figure}[!ht]
\centering
\includegraphics[trim=15mm 30mm 45mm 15mm, clip,width=3.5in]{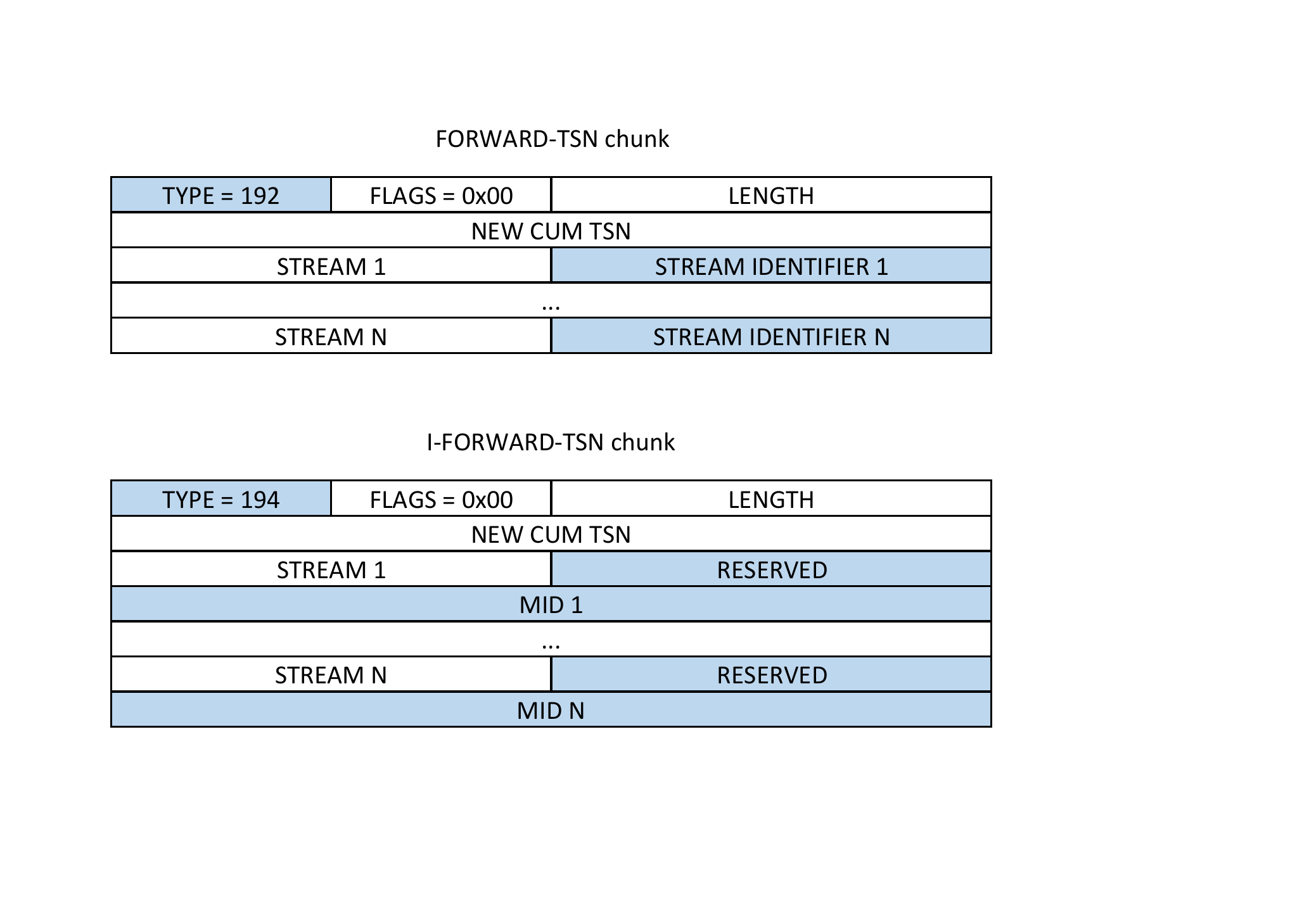}
\caption{FORWARD-TSN vs I-FORWARD-TSN chunk structure}
\label{fig:chunk-iforward}
\end{figure}

Furthermore, when using the partial reliability extension~\cite{rfc3758}, the FORWARD-TSN chunk contains the SSN field.
When using this protocol extension in combination with user message interleaving, an I-FORWARD-TSN chunk has to be used to account for the 32-bit MID.
The I-FORWARD-TSN chunk is shown in Figure~\ref{fig:chunk-iforward}. 

\section{Stream Scheduler}
\label{sec:stream-scheduler}
\begin{figure}[!ht]
\centering
\includegraphics[trim=5mm 25mm 5mm 20mm, clip,width=3.5in]{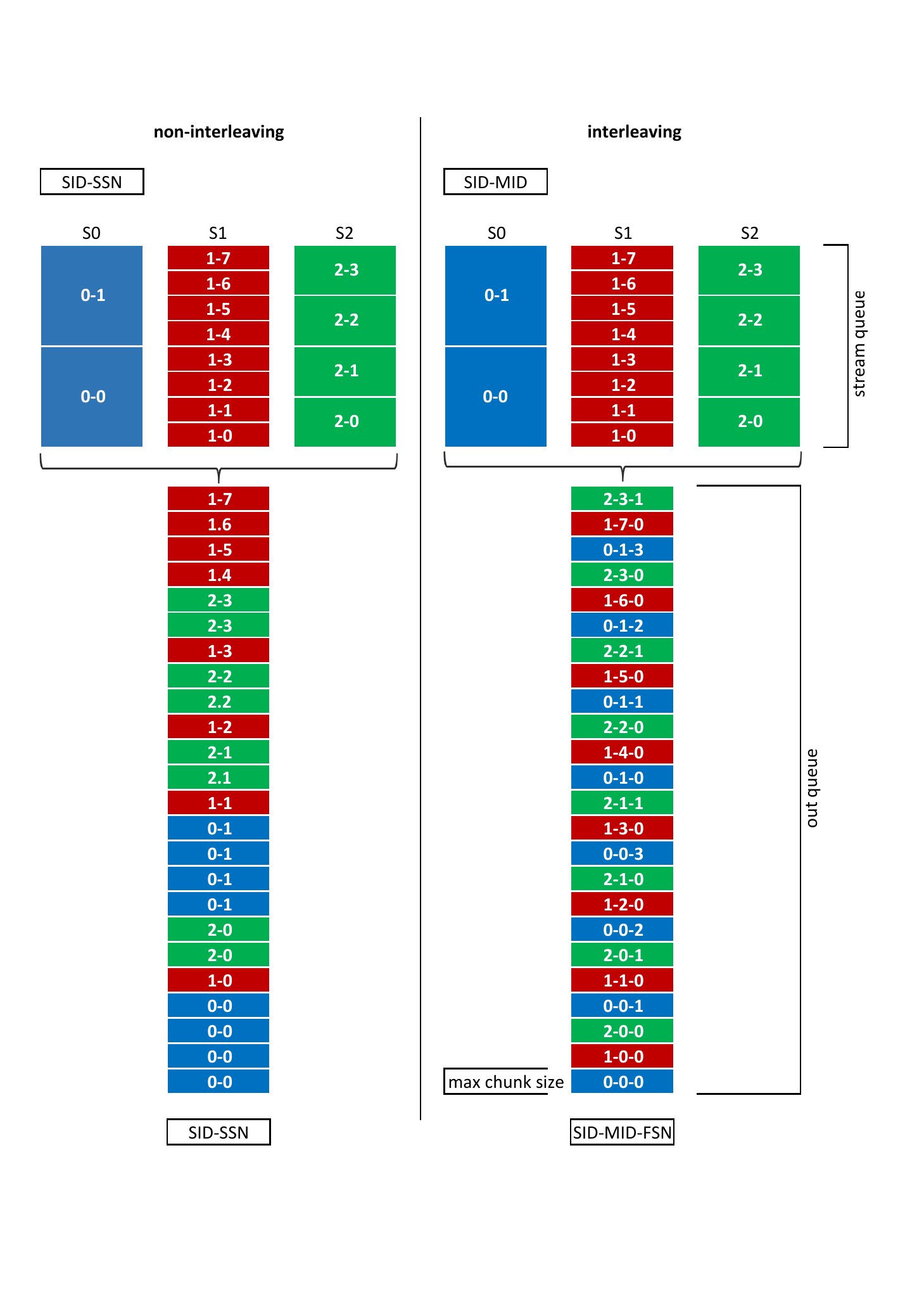}
\caption{Message queueing with and without interleaving by a round robin scheduler}
\label{fig:interleaving}
\end{figure}

Every SCTP association has an out queue and an additional queue for every stream as illustrated in Figure~\ref{fig:interleaving}.
User data from the application is stored into the corresponding stream queue.
The out queue contains the user messages which are in sending order and fragmented if necessary.
Stream schedulers are used to choose the next stream queue to put data into the out queue and are configurable by the application.
\cite{draft-ndata} not only specifies user message interleaving but also introduces a list of stream schedulers for SCTP. 
For each scheduler it is described how they operate when user message interleaving is used or not.
The stream schedulers without user message interleaving support are already implemented in the existing SCTP model of INET.

Figure~\ref{fig:interleaving} illustrates differences between an interleaving and non-interleaving scheduler by taking the round robin scheduler as an example.
When operating in non-interleaving mode the round robin scheduler chooses the first stream where the chunk zero has to be fragmented in four chunks to not exceed the maximum chunk size.
The scheduler keeps locked on the first stream (S0) until all four fragments of message zero are queued in the out queue before continuing with the next stream (S1).
This introduces the sender side head-of-line blocking.

In interleaving mode the scheduler chooses the first stream (S0) and iterates over all streams chunk by chunk regardless of fragmentation.

\section{Implementation}
\label{sec:implementation}
The implementation of the SCTP message interleaving extension has been done in two steps in the INET source code.

In a first step, we extended the SCTP model itself by adding the new chunk types, interleaving logic at the sender side and extended the existing stream schedulers with interleaving support.

To configure the interleaving extension we added the \texttt{iData} parameter to the SCTP model.
When the \texttt{iData} parameter is set to \texttt{true} the SCTP model offers message interleaving support in the 4-way-handshake to its peer.
If both peers offer interleaving support, the I-DATA and I-FORWARD-TSN chunks are used instead of the DATA and FORWARD-TSN chunks, and the SCTP model operates in interleaving mode.

Whenever receiving an I-DATA chunk SCTP iterates over the receiver queue to determine if it can successfully reassemble the data message by comparing MIDs and FSNs.

In the second step, we added support for the external interface and PCAP recorder files by adding the chunk types to the serializer and deserializer modules of INET.
We will address this topic in detail in the validation section. 

As mentioned in Section \ref{sec:stream-scheduler}, the SCTP model already includes all stream schedulers except the weighted fair queueing scheduler, but all schedulers lack the interleaving support.

\section{Validation}
\label{sec:validation}
To ensure the correct implementation of the interleaving extension, we used multiple techniques to validate the model operation within the INET framework (3.4) of the OMNet++ (5.0)
simulation environment. 

\subsection{Manual Packet Flow Inspection using Wireshark}
We were able to use Wireshark (2.0.2) to analyze the recorded PCAP files, since we implemented support for the SCTP interleaving extension earlier.
This helped us a lot to validate the correct message flow and find bugs in our implementation.
\subsection{Interoperability Testing}
After we successfully validated the model with Wireshark we ran tests with the SCTP implementation of the FreeBSD (11-current) kernel via the external interface.
We focused on the interleaving negotiation in the 4-way-handshake and the data transfer using I-DATA chunks.
Bulk transfer with multiple message loss rates was used and no issues found.

\subsection{Automated Packet Flow Inspection using Packetdrill}
Packetdrill\cite{packetdrill} is a script-based testing tool for transport protocols on Unix-based operating systems which has been released in 2013 by Google.
The great advantage of Packetdrill is the testing of specific scenarios like the usage of wrong parameters, which is not possible in usual test setups.
To benefit from this advantage, we ported Packetdrill to INET \cite{inet-packetdrill}.
It is already part of the current version and supports UDP, TCP and SCTP.

To be able to test the I-DATA chunk, Packetdrill had to be extended by the new chunk type, the new parameters and code to create new chunks to compare  them to the ones sent by the SCTP stack. 
As the interleaving support is negotiated in the handshake, the new socket options had to be added to trigger the \texttt{iData} parameter mentioned in section~\ref{sec:implementation}.
Our goal to run the same test scripts as the kernel implementation was achieved by porting the use of \texttt{ifdef} clauses from the kernel version.
Thus, the script can be tailored to handle OS specific features.
An example are error causes, that are not part of the simulation.

As the SCTP I-DATA feature was already integrated in Packetdrill for kernel SCTP, there were about 100 tests that could be applied to INET.
Several of those tests focus on the appropriate handling of received packets resulting from an inappropriate behavior of the sender.
While testing this for a real implementation is important, simulation might assume appropriate behavior of the sender.
Therefore, these test would not be passed by the simulation and have been disabled.
Nevertheless, more than 80 tests were passed that focussed on the conformance with the protocol specification.
These tests cover the negotiation of the extensions, the correct handling of the TSNs, and sending and receiving of fragmented messages.

\subsection{Measurements}
\label{sec:measurements}
To measure the improvements by message interleaving we built a typical WebRTC data channel scenario with two competing streams within one association.

\begin{figure}[!ht]
\centering
\includegraphics[trim=0mm 5mm 0mm 5mm, clip,width=3.0in]{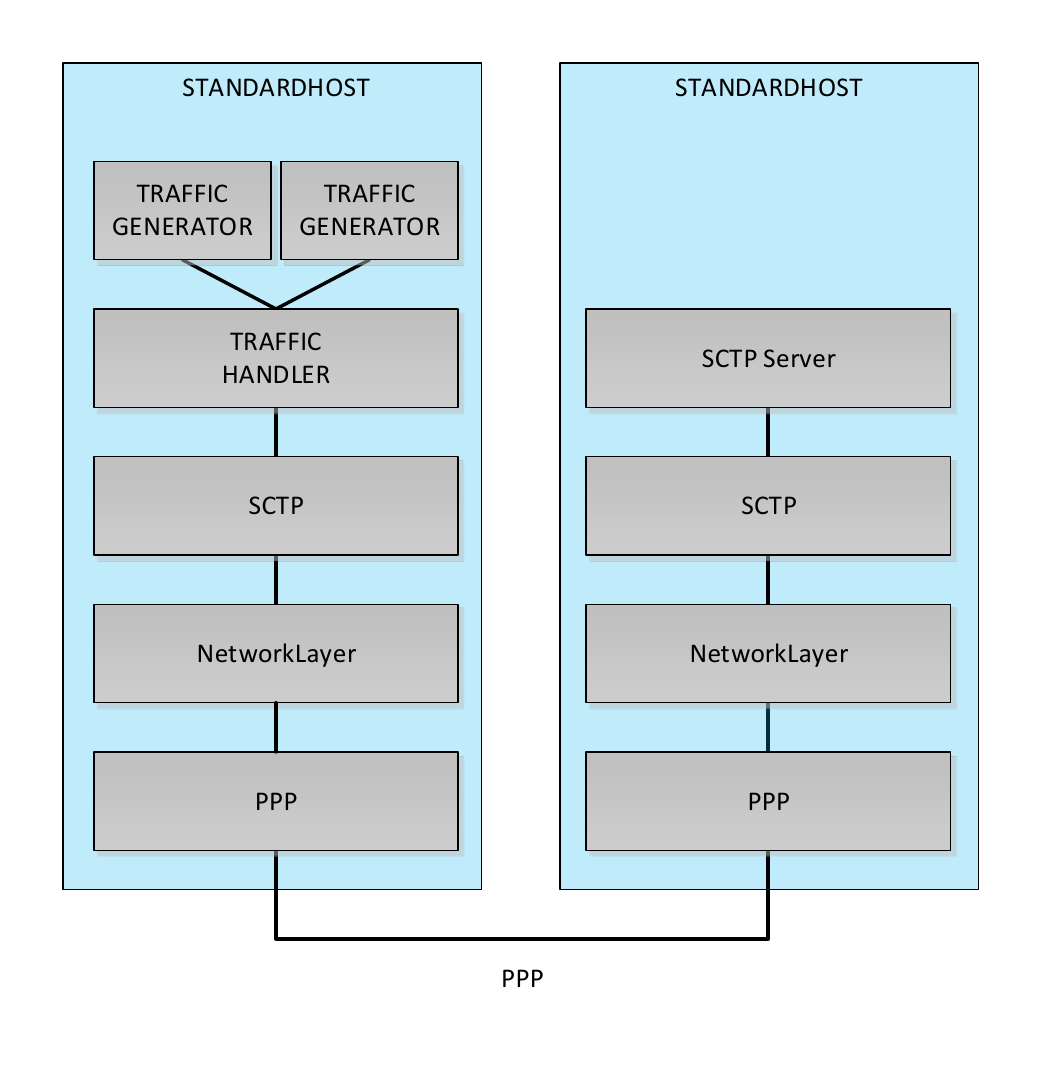}
\caption{Traffic generator and handler architecture}
\label{fig:measure-architecture}
\end{figure}

To have an adjustable testbed, we created a traffic generator module and a traffic handler module as shown in Figure~\ref{fig:measure-architecture}.
Each generator represents a single SCTP stream and is connected to a handler which establishes the connection to a remote peer.

While the first stream has a low priority and is saturated by large messages the second stream sends small messages with a high priority.
This is a common scenario for WebRTC applications like a chat program with file transfer capabilities.

While keeping the small messages between 8 - 16 bytes we increased the message size of the larger messages from 4 kilobytes to 128 kilobytes.
We were mainly interested in the message end-to-end delay.

\begin{figure}[!ht]
\centering
\includegraphics[trim=0mm 5mm 0mm 5mm, clip,width=2.5in]{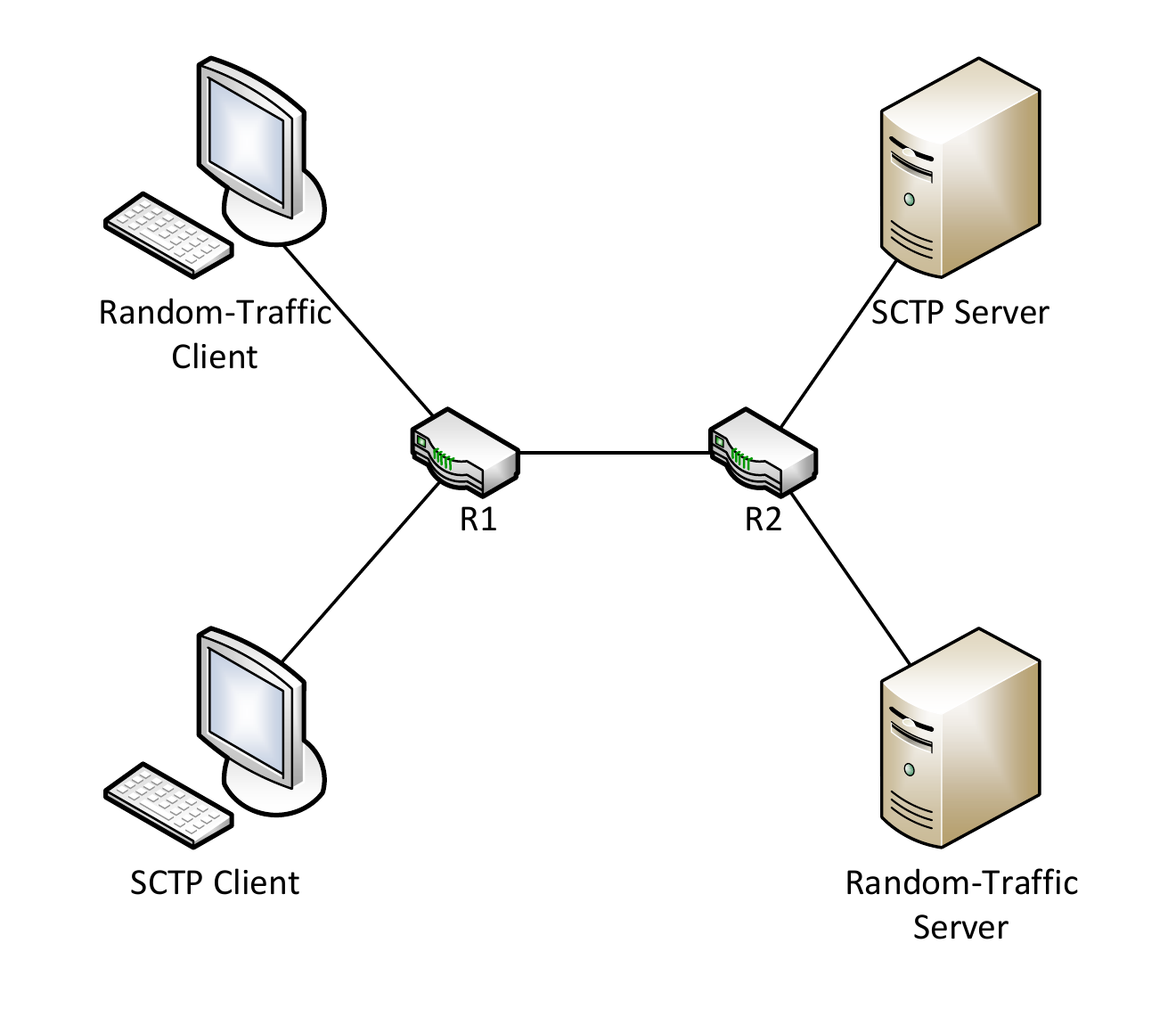}
\caption{Bottleneck scenario with SCTP and background traffic}
\label{fig:scenario}
\end{figure}

The traffic generators are intentionally separated from the traffic handler to support different generators with the same handler or the same generator for several protocols.
Figure \ref{lst:parameter} shows the configuration of a simple traffic generator as a saturated sender with an increasing message size for every run.
The traffic handler coordinates the generators and communicates with the SCTP module.
On the receiver side, we modified the SCTPServer to record the message delay not only per association but also for every stream.

\begin{figure}[t]
\begin{lstlisting}
# settings for client - first generator
...gen[0].typename = "TrafficgenSimple"
...gen[0].name = "low prio"
...gen[0].id = 1
...gen[0].priority = 128
...gen[0].packetCount = -1
...gen[0].packetSize = ${ps=4 .. 128 step 2}kB
...gen[0].packetInterval = 0ms
...gen[0].startTime = 5s
...gen[0].stopTime = 65s
\end{lstlisting}
\caption{Settings for a saturated traffic generator}
\label{lst:parameter}
\end{figure}

To achieve a more realistic testing scenario and get some randomness, we used a scenario where the SCTP traffic competed with a random UDP sender which sent a small amount of traffic - see Figure~\ref{fig:scenario}.

While increasing the message size of the saturated low priority stream we expected a linearly increasing delay for the high priority stream in non-interleaving mode.

\begin{figure}[!ht]
\centering
\includegraphics[trim=5mm 0mm 5mm 0mm, clip,width=3.5in]{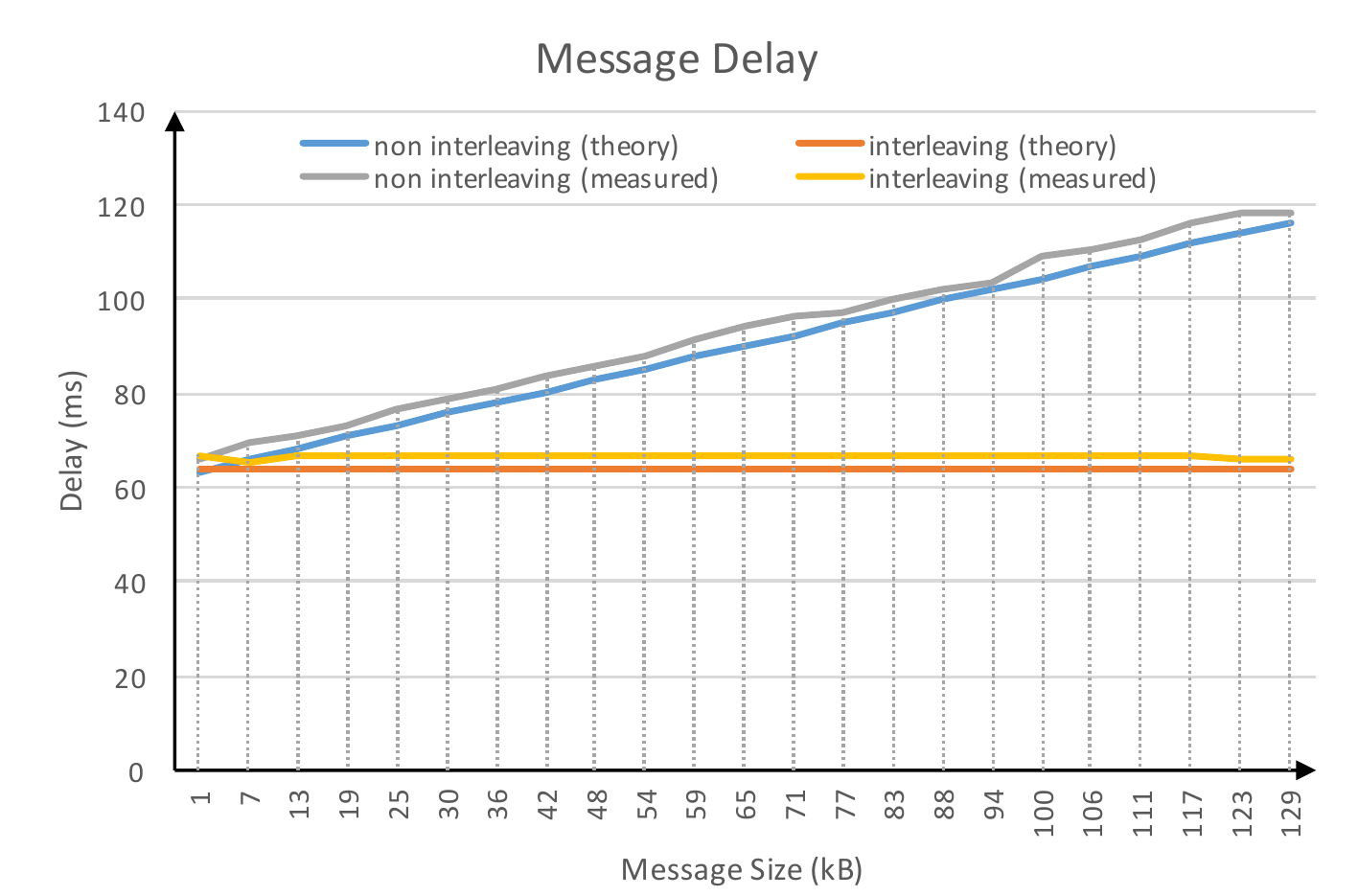}
\caption{Message delay with and without interleaving}
\label{fig:measurement}
\end{figure}

For the interleaving mode scenario, we expected the high priority stream delay at a constant level.
As shown in Figure~\ref{fig:measurement}, the delay for non-interleaving mode rises constantly while staying on a constant level when operating with interleaving.

The theoretical message delay \(d_n (s_\text{msg})\) in the non-interleaving mode and
\(d_i (s_\text{msg})\) in the interleaving mode depends on the user message size \(s_\text{msg}\)
and is calculated by adding
the link delay \(d_\text{link}\),
the delay caused by the link buffers \(d_\text{buffer}\) and
the specific delay caused by the SCTP buffers which is different for interleaving and non-interleaving mode.

For computing the specific delay in non-interleaving mode,
we take the user message size \(s_\text{msg}\) and calculate how many
SCTP packets are required to transfer the user message by dividing the user message
size by the maximum fragment size \(s_\text{frag}\).
The overhead is the product of this number and the corresponding header
size \(s_\text{hdr}\) for SCTP, IP and PPP. Adding the user message size to the overhead and
dividing by two gives the average buffered head-of-line-blocking size.
Finally dividing by the available bandwidth \(\text{bw}\) gives the specific delay in the
non-interleaving mode.
For the interleaving mode the specific delay is calculated by dividing the maximum transmission
unit \(\text{mtu}\), divided by 2, by the link bandwidth.

\begin{eqnarray*}
    d_n(s_\text{msg}) = \frac{s_\text{msg} + s_\text{hdr} \cdot \lceil\frac{s_\text{msg}}{s_\text{frag}}\rceil}{2 \cdot \text{bw}}
     + d_\text{link} + d_\text{buffer}
\end{eqnarray*}

\begin{eqnarray*}
    d_i(s_\text{msg}) = \frac{\text{mtu}}{2 \cdot \text{bw}} + d_\text{link} + d_\text{buffer}
\end{eqnarray*}

Figure~\ref{fig:measurement} shows a comparison between our expected results and the results from our simulation which meets our expectations.

\section{Conclusion and Outlook}
The message interleaving extension for SCTP solves head-of-line-blocking for fragmented messages in SCTP.
We integrated the extension into the SCTP model, verified its correct implementation by internal tests with Packetdrill as well as with interoperability tests with the real implementation in FreeBSD.
In our ongoing work, we will focus on buffer optimization and additional stream schedulers like the weighted fair queueing scheduler.
We will contribute the interleaving extension and the extended stream scheduler to the INET framework.

\section{Acknowledgements}
We would like to thank Julius Flohr from the University of Duisburg-Essen for the cooperation on the generator module.

\bibliographystyle{IEEEtran}
\bibliography{paper}

\end{document}